\documentclass[%
 reprint,
 amsmath,amssymb,
superscriptaddress
]{revtex4-1}
\usepackage{xparse}

\usepackage{appendix}
\usepackage{graphicx}% Include figure files
\usepackage{dcolumn}% Align table columns on decimal point
\usepackage{bm}% bold math
\newcommand{\ket}[1]{\ensuremath{\left| #1 \right>}}

\newcommand{\matrixel}[3]{\ensuremath{\left< #1 \vphantom{#3} \right| #2 
\left| #3 \vphantom{#1} \right>}}

\usepackage{bbold}
\begin{document}
\title{Dynamical Quantum Phase transitions and Recurrences in the Non-Equilibrium BCS model}% Force line breaks with \\

\author{Colin Rylands}
 \email{crylands@umd.edu}
\affiliation{Joint Quantum Institute and
 Condensed Matter Theory Center, University of Maryland, College Park, MD 20742, USA}

\author{Victor Galitski}
\affiliation{Joint Quantum Institute and
	Condensed Matter Theory Center, University of Maryland, College Park, MD 20742, USA}

\date{
    \today
}

\date{\today}
\begin{abstract}
Non-equilibrium aspects of the BCS model have fascinated physicists for decades, from the seminal works of Eliashberg to modern realizations in cold atom experiments. The latter scenarios have lead to a great deal of interest in the quench dynamics of fermions with pairing interactions. The recently introduced notion of a dynamical quantum phase transition is an attempt to classify the myriad of possible phenomena which can result in such far from equilibrium systems. 
These are defined as non-analytic points of the logarithm of the Loschmidt echo and are linked to oscillations in the dynamics a systems order parameter. In this work we analytically investigate the relation between DQPTs and oscillation of the superconducting order parameter in quenches of the BCS model. We find that each oscillation of the order parameter is accompanied by a DQPT which is first order in nature. We show this for a variety of initial states and furthermore find that when the order parameter  attains a constant steady state then no DQPTS occur. 
\end{abstract}

\maketitle
\section{Introduction}
Equilibrium phase transitions (EPTs), both classical and quantum are quite well understood. This is grounded in the existence of a ubiquitous object, the partition function which characterizes the system. In the thermodynamic limit, the  logarithm of the partition function, the free energy may exhibit non analyticities as a function of temperature or some other system parameter, signifying a phase transition. This behavior is then displayed by the systems observables which are given by derivatives of the free energy and so maybe non analytic themselves. Combined with the notion of universality and the renormalization group\cite{WilsonRMP} equilibrium phases and the subsequent properties of many disparate systems can be efficiently studied.

Away from equilibrium the situation is less straightforward. For  closed quantum systems which are far from equilibrium the natural object to study is the time evolution operator, $U(t)$. This object however, can defy calculation even in non-interacting systems as it depends on both the Hamiltonian of the system and how it was taken out of equilibrium. The simplest type of non equilibrium scenario one can study is the sudden quantum quench\cite{PolRev, MitraRev, CazalillaJstat,CalabreseCardyJstat, CauxJstat, EsslerFagottiJstat, RylandsAndreiARCMP}. A system is initially prepared in a state, $\ket{\Psi_i}$ which is taken to be an  eigenstate of some Hamiltonian, $H_i(\lambda_i)$ which depends upon a parameter $\lambda_i$. This is then rapidly changed, $\lambda_i\to\lambda_f$ and the system allowed to evolve according to $H(\lambda_f)$. Such quenched systems are routinely created experimentally in a multitude of platforms most notably in cold atom gas systems\cite{IBRMP}. The theoretical simplicity and experimental relevance of the sudden quantum quench has resulted in this being by far the most widely studied type of non-equilibrium system. 

In this context it was proposed that a simpler quantity to study is the Loschmidt amplitude\cite{HeylPolKeh, Heylreview}
\begin{eqnarray}
G(t)&=&\matrixel{\Psi_i}{U(t)}{\Psi_i}=\matrixel{\Psi_i}{e^{-iH(\lambda_f)t}}{\Psi_i},
\end{eqnarray}
which is the expectation value of the time evolution operator in the initial state. $G(t)$ resembles a boundary partition function and likewise its logarithm may exhibit non analytic points as a function of $t$\cite{CardCal}. By analogy with EPTs these are called dynamical quantum phase transitions (DQPTs). Unlike the equilibrium system however DQPTs may occur for finite size systems and periodically in time. We denote the period by $t_\text{DQPT}$. As a function of $z\in \mathbb{C}$, $G(z)$ may have lines of zeros throughout the complex plane. EPTs occur when these zeros cross the imaginary axis\cite{Fisher65} while DQPTs occur when they cross the real axis. If there are no non-analytic points within the upper right quadrant of the complex plane then one can analytically continue $G(t)$ to imaginary time and study the dynamics of the system using standard techniques of equilibrium systems\cite{WilsonRMP, CardyBCFT}. The appearance of non analytic points in  the upper quadrant signals the breakdown of this method and furthermore the breakdown of any real time perturbative methods once they cross the real axis, i.e when a DQPT occurs.

 A related quantity of interest is the Loschmidt echo, $L(t)=|G(t)|^2$ which is the probability that the system returns to its initial state after a time $t$.   The echo shares the same analytic properties as $L(t)$ and DQPTs can likewise be defined as non-analytic behaviour of $\log{L(t)}$.
The relevance of $G(t)$ or $L(t)$ for the dynamics of the system and in particular the behaviour of its observables as a function of time, is less obvious than in the equilibrium case. Indeed, derivatives of $\log{G(t)}$ do not give expectation values of observables. 
It has been shown however, notably in the Ising model and its variants \cite{HeylPolKeh} and more recently in the Bose Hubbard model\cite{LackiHeyl} that the period of oscillations of the order parameter coincides exactly with $t_\text{DQPT}$ in certain quenches. The connection between these two quantities is through the amount of work, $W$, done on the system during the quench and via analogy with zero temperature EPTs or quantum phase transitions\cite{HeylPolKeh, Heylreview, Heylsurvey}. In this regard $W$ plays the role that temperature does in EPTs. The post quench behaviour of a system will depend upon the amount of energy it has absorbed during the quench and so its observables should depend upon $W$\cite{Goold}. By definition  $L(t)$ is the probability that no work be done on the system and so one can expect the behaviour of $L(t)$ to be related to that of the observables if $W=0$. By analogy with quantum phase transitions which happen at zero temperature but affect properties of the system at $T>0$\cite{Sachdev}, the behaviour of the $L(t)$ can be expected to affect the system properties even when $W>0$.

DQPTs have by now been investigated in numerous systems and for various quenches and although there exist a number of analytical studies, results are predominantly based upon numerical work and for finite size systems \cite{ poz1, poz2,poz3,PerfettoPiroliGambassi, RylandsMTM, RylandsAndreiLLwork, TrapinHeyl, KhatunBhattacharjee,VajnaBalazs, LangFrankHal,HeylScaling,AndSirk, SharmaSuzukiDutta, KarraschSchuricht, KennesSchurichtKarrasch, Fogarty, VoskAltman, GurarieDQPTS, HalimehYegGurar, BudichHeyl, HeylBudich, BhattacharyaBandDutta, BhattacharyaDutta, CanoviWernerEck}.
In particular we are not aware of any studies in which both the dynamics of the order parameter and the Loschmidt echo are calculated analytically and done so for a range of initial states. The aim of the present study is to fill this gap. Ideally we would like to study an experimentally relevant model which exhibits nontrivial quench dynamics, supports DQPTs and admits an analytical description of both its order parameter and the Loschmidt amplitude when quenched from a variety of initial states. Fortunately, such desirable qualities are possessed by the reduced BCS model (see \eqref{H}), hereafter simply refereed to as the BCS model. 

The BCS model is one of the most comprehensively studied models of non-equilibrium physics. Originally investigated in the context of solid state systems through the seminal works of Eliashberg and collaborators\cite{Eliashberg1, Eliashberg2, Eliashberg3}, the model has garnered much attention in the field of  ultra cold atom experiments\cite{Bloch}. In these experiments, closed systems of fermions with tunable pairing interactions can be readily created. The ability to vary system parameters in real time within these experimental systems allows one to study  the quench dynamics of a closed quantum system. 
There exists an extensive literature on the quench dynamics of the BCS and related models detailing its behaviour from a variety of initial states and changes of system parameters\cite{BarankovLevitov,BarakovSpivakLevitov, BarankovLevitov2,  Gurarie, YuzbashDzero, YuzbashAltKuzEnol,  Yuzbash, YuzbashAltKuzEnol, DzeroyuzbashAltColeman, YuzbashKuzAlt}. Broadly, the dynamics of its order parameter post quench can be classified into 2 distinct classes depending on the choice of initial state and final Hamiltonian parameters. In the first, the order parameter, $\Delta(t)$ undergoes persistent  oscillations while in the second the oscillations decay and the system is characterized at long time by a constant $\Delta(t)=\Delta_\infty\ge 0$. Relying upon this previous work we show that in the former scenario each oscillation is accompanied by a DQPT which occurs exactly when  $\Delta(t)$ reaches a maximum. These DQPTs are distinct from those discovered previously in that they occur only in the thermodynamic limit and are akin to first order EPTs i.e, $\partial_t\log{L(t)}$ is discontinuous. In the latter scenario when oscillations are absent, we find that there are no DQPTs at long time. Furthermore, we show that this connection between DQPTs and maxima of $\Delta(t)$  exists for many different choices of initial state and can be extended to encompass other related central spin\cite{YuzbashAltKuzEnol} and cold atom models \cite{YuzbashKuzAlt}.

\section{Hamiltonian and Quench dynamics} The Hamiltonian of the BCS model is given by
\begin{eqnarray}\label{H}
H=\sum_{p,\sigma}\epsilon_pc^\dag_{p,\sigma}c_{p,\sigma}-\lambda\sum_{p,q}c^\dag_{p,\uparrow}c^\dag_{-p,\downarrow}c_{q,\uparrow}c_{-q,\downarrow}
\end{eqnarray}
where $c^\dag_{k,\sigma}, c_{k,\sigma}$ are creation and annihilation operators for fermions with spin $\sigma=\uparrow, \downarrow$ and momentum $k$. They have single particle energy levels  $\epsilon_k$ and interact via a pairing interaction of strength $\lambda$. The model separates into sectors wherein each level is singly occupied, called the blocked sector, or either empty or doubly occupied called the unblocked sector.  We shall consider here the case where there are no states in the blocked sector, levels are either empty or doubly occupied and are interested in the quench where we change the coupling constant from $\lambda_i\to\lambda_f$ at some time $t^*$ where $\lambda_i\ge 0$.

The Hamiltonian is quantum integrable\cite{RICHARDSON, Gaudin} and can be solved via Bethe Ansatz however since it contains infinite range interactions the mean field description becomes exact in the thermodynamic limit and provides a simpler approach to the system. 
This remains true even out of equilibrium and so we have that at time $t$ the system is in the state
\begin{eqnarray}\label{psit}
\ket{\Psi_i(t)}=\prod_p\left[u_p(t)+v_p(t)c^\dag_{p,\uparrow}c^\dag_{-p,\downarrow}\right]\ket{0}. 
\end{eqnarray}
Here $\ket{0}$ is the vacuum which contains no particles. The coefficients $u_p(t)$ and $v_p(t)$ are solutions of the time dependent Bogoliubov-de-Gennes (BdG) equations,
\begin{eqnarray}\label{BdG}
i\partial_t\begin{pmatrix}u_p(t)\\
v_p(t)
\end{pmatrix}=\begin{pmatrix} \epsilon_p& \Delta(t)\\
\Delta(t)& -\epsilon_p
\end{pmatrix}\begin{pmatrix}u_p(t)\\
v_p(t)
\end{pmatrix},
\end{eqnarray}
where $\Delta(t)$ is the time dependent superconducting order parameter. It is defined as 
\begin{eqnarray}\label{SCcondition}
\Delta(t)&=&\lambda_f \sum_p\matrixel{\Psi_i(t)}{c^\dag_{p,\uparrow}c^\dag_{-p,\downarrow}}{\Psi_i(t)},\\\label{Delta}
&=&\lambda_f\sum_{p}u_p(t)v_p^*(t).
\end{eqnarray}
The simplicity of the state \eqref{psit} means that the Loschmidt amplitude can be readily evaluated, 
\begin{eqnarray}\label{Gt}
G(t)=\prod_p\left[u_p^*(0)u_p(t)+v_p^*(0)v_p(t)\right]
\end{eqnarray}
where the initial state encoded in the initial conditions $u_p(0), v_p(0)$. 
The solutions of  \eqref{BdG} subject to the self consistency condition \eqref{Delta} constitute an exact description of the quench dynamics of the system. This represents a dramatic simplification of the full many body problem, however the solution of \eqref{BdG} for arbitrary $\Delta(t)$  is not known in closed form. Despite this there exists a number of exact, non trivial solutions to the self consistent problem which also correspond to physically relevant quenches. 

These exact solutions where discovered first in studies of Peierls instabilities in one dimensional conductors\cite{TakayamaLinLiuMaki,Brazovskii, HorovitzComment, Horovitz, BrazovskiiKirovaMatveenko}. It was shown that such materials could could support solitonic excitations of its order parameter and moreover periodic spatial superstructures. The equations governing these soliton lattices can be ampped to \eqref{BdG} and \eqref{Delta} with space exchange for time. Therefore the quenched BCS model can support such solitonic like behavior in the dynamics of $\Delta(t)$ and in particular can exhibit persistent oscillations in time. 

Other classes of solutions were subsequently constructed in which $\Delta(t)$ exhibited damped or overdamped oscillations and attained a steady state value at long time. In the following sections we review the construction of these solutions as well as calculate the Loschmidt echo. By expressing the echo directly in terms the order parameter we construct a ``dynamical free energy"\cite{Fagotti} and can relate the DQPTs directly to $\Delta(t)$. 

\section{Persistent Oscillations and DQPTs}

\subsection{A single revival}
The simplest quench problem is to take the system to be initially a  free Fermi gas described by \eqref{H} with $\lambda_i=0$. Moreover we consider the case when $\ket{\Psi_i}$ is the ground state with $u_p(0)=\Theta(\epsilon_p),v_p(t)=\Theta(-\epsilon_p)$. 
Following [\citenum{BarakovSpivakLevitov}]  the quench dynamics are solved by  rewriting the BdG equations in terms of  $\omega_p=u_p/v_p$ and $f_p=u_pv_p$ for $\epsilon_p>0$ ($\epsilon_p<0$ is similar due to particle hole symmetry but with the solutions  $u_p\leftrightarrow v_p$ and $\epsilon_p\to |\epsilon_p|$). These quantities satisfy the following equations 
\begin{eqnarray}\label{omega}
i\partial_t\omega_p(t)=2\epsilon_p \omega_p+\Delta(t)(1-\omega_p^2)\\\label{f}
i\partial_tf_p(t)=\Delta(t)(\omega_p+\omega_p^{-1})f_p.
\end{eqnarray}
The solution for $\Delta(t)$ follows from the substitution 
\begin{eqnarray}
\omega_p(t)=\frac{1}{\Delta(t)}\left[2\epsilon_p+i\partial_t(\log{(\Delta(t))})\right]\\
f_{p}(t)=Ne^{-i\int^t\mathrm{d}s\Delta(s)\left[\omega_p(s)+\omega^{-1}_p(s)\right]}
\end{eqnarray}
where $N$ is some normalisation. The order parameter is given by
\begin{eqnarray}\label{Deltasingle}
\Delta(t)=\frac{\Delta_f}{\cosh{\left(\Delta_f(t-t_0)\right)}}
\end{eqnarray}
where $t_0$ an integration constant and $\Delta_f$ determined by the self consistency condition \eqref{SCcondition}\footnote{The ground state of free Fermi gas is actually an eigenstate of BCS Hamiltonian and so is an unstable fixed point of the dynamics. In principle therefore to peak will occur only at $t_0\to\infty$. Alternatively one can understand the free the initial state as being the limit of having an infinitesimal $\lambda_i$. This agrees with the $\Delta_-\to 0$ limit of \eqref{Deltamulti}}. The order parameter therefore exhibits a single collapse and revival peak with the maximum occurring at $t=t_0$.

In order to calculate $G(t)$  we require  $u_p$ and $v_p$ which can be determined by writing $u^2_p=\omega_p(t)f_p(t)$. After substituting in for $\omega_p,f_p$ and carrying out the integration we get 
\begin{eqnarray}\label{u}
u_p(t)=\left[1-\frac{\Delta^2(t)}{(2\epsilon_p)^2+\Delta_f^2}\right]^\frac{1}{2}e^{-i\arctan{\left(\frac{\sqrt{\Delta_f^2-\Delta^2(t)}}{2\epsilon_p}\right)}+i\phi}
\end{eqnarray}
where
$\phi=\arctan{\left(\frac{\Delta_f}{2\epsilon_p}\right)}-\epsilon_pt$ along with $v_p(t)=u_p(t)\omega_p(t)$. 
 We can check that these give a consistent solution by computing $u_pv^*_p$
 \begin{eqnarray}
u_pv^*_p&=&\left[\frac{\Delta(t)}{(2\epsilon_p)^2+\Delta_+^2}\right](2\epsilon_p-i\sqrt{\Delta_+^2-\Delta^2(t)})
 \end{eqnarray}
 Inserting this into \eqref{SCcondition} as well as the corresponding expressions for $\epsilon_p<0$ it is found that $\Delta_f=2De^{-\frac{1}{\nu\lambda_f}}$ where $\nu$ is the density of states around the Fermi energy and $D$ is the half bandwidth of the system. This is the gap that would be present in equilibrium with $\lambda=\lambda_f$. The Loschmidt echo is now straightforward to evaluate. Using \eqref{Gt} and \eqref{u} we find 
\begin{eqnarray}
L(t)&=&\prod_p\left[1-\frac{\Delta^2(t)}{(2\epsilon_p)^2+\Delta_f^2}\right]\\
&=&\exp{\left(\nu\int_0^D\mathrm{d}\epsilon\,\log{\left[1-\frac{\Delta^2(t)}{(2\epsilon)^2+\Delta_f^2}\right]}\right)}.
\end{eqnarray}
 Where in going to the second line we have gone to the continuum description and assumed a flat density of states  Evaluating the integral for large $D\gg \Delta_f$ we obtain the compact result
\begin{eqnarray}\label{L}
L(t)=e^{-\pi\nu\left[1-\sqrt{1-\frac{\Delta(t)^2}{\Delta_f^2}}\right]}.
\end{eqnarray}
This expression explicitly shows the connection between the behaviour of the order parameter and the Loschmidt echo thereby connecting revivals with DQPTs. We see that at the peak of the order parameter, i.e. $t=t_0$ where $\Delta(t_0)=\Delta_f$ the echo becomes non-analytic.  Thus, the Loschmidt echo is analytic as the order parameter approaches its equilibrium value however upon attaining this a DQPT occurs which coincides with the order parameter displaying non-equilibrium behaviour i.e. turning away  from its equilibrium value,  $\Delta_f$.

We note that upon analysing the exact eigenstates of $H$ for a finite system\cite{RICHARDSON} as well as their form factors\cite{Slavnov, KorepinBook,Kitanine,Claeys} the DQPT occurs only in the thermodynamic limit wherein the mean field analysis is valid. This behaviour is distinct from that which occurs in the Ising model which exhibit DQPTs regardless of system size.

The quantity  
\begin{eqnarray}\label{l}
l(t)\equiv -\log{L(t)}=\pi\nu\Delta_f\left[1-\sqrt{1-\frac{\Delta(t)^2}{\Delta_f^2}}\right]
\end{eqnarray} 
can be viewed as a dynamical free energy. Expanding around the DQPT point we have that $l=\pi\nu\Delta_f(1-|t-t_0|)$ which shows the nature of the singular point. The discontinuity in the first derivative is akin to a first order EPT, further marking this behaviour as distinct from the DQPTs in the Ising model.  Recently there have been attempts to construct Ginzburg-Landau type theories describing the system in the region of DQPTs and analyse their scaling behaviour\cite{HeylScaling,TrapinHeyl, KhatunBhattacharjee}. This has been achieved in certain quenches of the Ising model using a mapping to the equilibrium scaling RG flow of the model. \eqref{l} is an exact expression for the dynamical free energy which is valid at arbitrary time and, as we shall show in the next section, for a wide range of initial states. 

Before proceeding, it is instructive to compare \eqref{L} to the thermodynamic case obtained from the boundary partition function at inverse temperature $\beta=-it$. This was calculated in [\citenum{Galitski}] where it was shown that, rather surprisingly, the result is independent of $\Delta(\beta)$ and agrees with that of free fermions. The difference between the boundary free energy computed therein and $l(t)$ suggests an involved analytic structure of $G(z)$ in the complex plane.  

\subsection{Alternative initial states}
Although the previous analysis was carried out for $\ket{\Psi_i}$ being the free ground state it can be carried over to other initial states also. To do so one can note that the BdG equations take the form of the Schr\"{o}dinger equation for a driven two-level system and accordingly the exact time evolution operator can be constructed if two independent solutions can be found\cite{GangopadhyayDzeroGalitski}. By exploiting the particle-hole symmetry a second solution can be constructed from \eqref{u}. In general we have that $U(t)=\prod_pU_p(t)$ where 
\begin{eqnarray}\label{U}
U_p(t)=\begin{pmatrix}u_p(t)&v^*_p(t)\\
v_p(t)&-u^*_p(t)
\end{pmatrix}\begin{pmatrix}u_p(0)^*&v^*_p(0)\\
v_p(0)&-u_p(0)
\end{pmatrix}
\end{eqnarray}
for $\epsilon_p>0$ with $u_p,v_p$ given by \eqref{u} and a similar expression for $\epsilon_p<0$ with $u_p\leftrightarrow v_p$ and $\epsilon_p\to |\epsilon_p|$. In this notation the free ground state corresponds to $\ket{\Psi_i}=\left[\otimes_{\epsilon_p<0} (0,1)_p\right]\left[\otimes_{\epsilon_p>0} (1,0)_p\right]$. 
We may therefore consider alternative initial states, in particular  we may take excited states of the free (but still unblocked) gas. Particle like excitations are created by replacing $(1,0)\rightarrow(0,1)$ for some level, $\epsilon_p>0$, in the ground state. 
 In order for this to satisfy the consistency condition however we need to also create a hole-like excitation by making the opposite replacement  at the corresponding negative energy, $-\epsilon_p$. Thus we can move any number of cooper pairs from below the Fermi surface to the unoccupied mirror image above it without disturbing $G(t)$. This necessitates modification to the consistency condition. For example if we have these particle hole excitations in the initial state for energies $0<W\ll D$ we have instead $\Delta'_f\approx\Delta_f-2W$
which reduces to the regular equation when $W\to 0$. The Loschmidt echo starting from this initial state takes the same form as before \eqref{L} as the particle-hole excitation corresponds to $u_p(t)\to-u_p^*(t)$ for the particle part and likewise for the hole part. Additionally, there is the appropriate replacement of the $\Delta_f\to\Delta'_f$.  Therefore when quenching from excited states or the ground state of a free Fermi gas to the BCS Hamiltonian a single DQPT occurs which coincides exactly with the peak of the soliton-like behavior of the order parameter.

\subsection{Persistent Oscillations}
We examine now the case when the initial state is the ground state of the BCS model with $\lambda_i>0$. In this scenario it was shown that provided $\Delta_i/\Delta_f\leq e^{-\pi/2}$, $\Delta(t)$ exhibits persistent oscillations\cite{BarankovLevitov}
\begin{eqnarray}\label{Deltamulti}
\Delta(t)=\Delta_+\text{dn}\left(\Delta_+(t-t_0),1-\Delta^2_-/\Delta^2_+\right)
\end{eqnarray}
dn$(x,k)$ being the Jacobi elliptic function. The period of the oscillations is $2K[(1-\Delta^2_-/\Delta^2_+)/\Delta_+]\equiv t_\text{DQPT}$ with $K$ the complete elliptic integral of the first kind. $\Delta_\pm$ are constants which are determined through \eqref{Delta} and are the upper and lower bounds of $\Delta(t)$,  $\Delta_-\le \Delta(t)\le \Delta_+$. In this situation the expressions for $u_p,v_p$ are more complicated but nevertheless can be determined along similar lines of the decomposition in \eqref{omega} and \eqref{f}. Of particular interest is the case when
 $\Delta_-\ll \Delta_+$. In this regime the order parameter behaves as a sum of widely separated peaks $\Delta(t)\approx\sum_{n=1}^\infty\Delta_+\text{sech}{\left[\Delta_+(t- t_0+nt_\text{DQPT})\right]}$. The full expression for the echo then reduces to
\begin{eqnarray}
L(t)=e^{-\pi\nu\Delta_+\left[1-\sqrt{1-\frac{\Delta(t)^2}{\Delta_+^2}}\right]}
\end{eqnarray}
indicating the presence of a DQPT at each peak of the oscillation which reoccur with period $t_\text{DQPT}$. It can be checked that this also occurs for general $\Delta_\pm$.
\section{Absence of DQPTs in the absence of revivals}
The previous section detailed how an oscillatory order parameter is accompanied by a DQPT which occurs exactly when $\Delta(t)$ reaches a local maximum. We now turn to the case when the system is characterized at long time by a constant order parameter $\Delta(t)\to\Delta_\infty$, $\Delta_\infty\geq 0$. This occurs when quenching from the ground state of $H$ with with the ratio  $\Delta_i/\Delta_f\geq e^{-\pi/2}$. In this case the state of the system is determined by mapping the mean field system onto an equivalent classical spin system\cite{YuzbashDzero, YuzbashAltKuzEnol}. This auxiliary system is classically integrable and its dynamics solved exactly by means of the Lax vector formalism. In this manner the long time limit of $\ket{\Psi_i(t)}$ can be determined. Using the result derived in [\citenum{YuzbashDzero}] we find that the Loschmidt echo can be written
\begin{eqnarray}\label{Lness}
L(t)=\mathcal{F}\prod_{p}\left[1+\mathcal{K}(\epsilon_p)e^{-2it\sqrt{\epsilon_p^2+\Delta_\infty^2}}\right]
\end{eqnarray}
where $\mathcal{F}$ is a constant, independent of $t$ and $\mathcal{K}(\epsilon)$ is given by
\begin{eqnarray}
\mathcal{K}(\epsilon)=\frac{\frac{\Delta_i}{\sqrt{\epsilon^2+\Delta_i^2}}\sin{[\theta(\epsilon)/2}]}{1-\frac{\epsilon}{\sqrt{\epsilon^2+\Delta_i^2}}\cos{[\theta(\epsilon)/2}]}
\end{eqnarray}
with the angle of rotation, $\theta(\epsilon)$ determined through
\begin{eqnarray}
\sin^2[\theta(\epsilon)]=\frac{\mathcal{G}(\epsilon)}{2\pi^2}-\sqrt{\frac{\mathcal{G}^2(\epsilon)}{4\pi^4}-\frac{4\alpha \Delta^2_i}{\sqrt{\epsilon^2+\Delta^2_i}}}
\end{eqnarray}
with $\alpha=\log{(\Delta_i/\Delta_f)}$ and $\mathcal{G}(\epsilon)=\pi^2+4\alpha^2+4\text{arcsinh}^2(\epsilon/\Delta_i)+8\alpha\epsilon \text{arcsinh}(\epsilon/\Delta_i)/\sqrt{\epsilon^2+\Delta_i^2}$. Despite the complex form of $\mathcal{K}$ \eqref{Lness} takes the standard form of for a quench between two quadratic theories\cite{HeylPolKeh, RylandsMTM}. Written in this fashion it is transparent that for a DQPT to occur it we require  $\mathcal{K}(\epsilon_p)e^{-2it\sqrt{\epsilon_p^2+\Delta_\infty^2}}=-1$ however it can be easily checked that $|\mathcal{K}(\epsilon)|<1$. Therefore DQPTs cannot take place. 

\section{Conclusion}
We have examined the existence of dynamical quantum phase transitions and their relation to oscillatory behaviour of the order parameter in quenches of the BCS model. We have shown analytically that when the order parameter exhibits soliton-like behaviour a DQPT occurs exactly when $\Delta(t)$ reaches a local maximum. This coincides with the equilibrium value of the order parameter and so a DQPT occurs when the system displays the far from equilibrium of  evolving away from its equilibrium position. Unlike previously studied examples, these DQPTs are first order and occur only in the thermodynamic limit.
In contrast, if the system attains a long time steady state wherein the order parameter is constant, DQPTs are absent. 
 This behaviour was shown to occur for a wide variety of initial states including excited ones, thus providing a solid connection between the behavior of the observables of a system and the existence of DQPTs in agreement with earlier results in the Ising\cite{HeylPolKeh} and Bose-Hubbard models\cite{LackiHeyl}. Unlike previous works however our calculation is the first to treat the problem analytically. We conclude by noting that similar solitonic solutions exist for related models describing Bose-Fermi gases\cite{YuzbashKuzAlt} and central spin systems\cite{Gaudin}. We expect that similar behavior occurs therein also.
\acknowledgements{This work was supported by NSF DMR-1613029, US-ARO (contract No. W911NF1310172), DARPA DRINQS program (C.R.), DOE-BES (DESC0001911), and the Simons Foundation (V.G.).}
\bibliography{bib}
\end{document}